\documentclass[letterpaper]{article} 
\usepackage{aaai25}  
\usepackage{times}  
\usepackage{helvet}  
\usepackage{courier}  
\usepackage[hyphens]{url}  
\usepackage{graphicx} 
\usepackage{natbib}  
\usepackage{caption} 
\usepackage{algorithm}
\usepackage{algorithmic}

\usepackage{newfloat}
\usepackage{listings}
\DeclareCaptionStyle{ruled}{labelfont=normalfont,labelsep=colon,strut=off} 
\floatstyle{ruled}
\newfloat{listing}{tb}{lst}{}
\floatname{listing}{Listing}
\title{Experimental Evidence That AI-Managed Workers Tolerate Lower Pay Without Demotivation}
\author{
    Mengchen Dong\textsuperscript{\rm 1}\thanks{Corresponding author: dong@mpib-berlin.mpg.de},
    Levin Brinkmann\textsuperscript{\rm 1},
    Omar Sherif\textsuperscript{\rm 1},
    Shihan Wang\textsuperscript{\rm 2},
    Xinyu Zhang\textsuperscript{\rm 2},
    Jean-François Bonnefon\textsuperscript{\rm 3},
    Iyad Rahwan\textsuperscript{\rm 1}\\
}
\affiliations{
    \textsuperscript{\rm 1}Center for Humans and Machines, Max Planck Institute for Human Development, Berlin, Germany\\
    \textsuperscript{\rm 2}Department of Information and Computing Sciences, Utrecht University, Utrecht, the Netherlands\\
    \textsuperscript{\rm 3}Toulouse School of Economics, Centre National de la Recherche Scientifique (TSM-R), University of Toulouse Capitole, Toulouse, France\\

}

\usepackage{bibentry}

\begin{document}

 \maketitle

\begin{abstract}
Experimental evidence on worker responses to AI management remains mixed, partly due to limitations in experimental fidelity. We address these limitations with a customized workplace in the Minecraft platform, enabling high-resolution behavioral tracking of autonomous task execution, and ensuring that participants approach the task with well-formed expectations about their own competence. Workers (N = 382) completed repeated production tasks under either human, AI, or hybrid management. An AI manager trained on human-defined evaluation principles systematically assigned lower performance ratings and reduced wages by 40\%, without adverse effects on worker motivation and sense of fairness. These effects were driven by a muted emotional response to AI evaluation, compared to evaluation by a human. The very features that make AI appear impartial may also facilitate silent exploitation, by suppressing the social reactions that normally constrain extractive practices in human-managed work. 

\end{abstract}

%

\section{Introduction}

Companies and organizations increasingly use AI-powered algorithms to manage human workers \cite{dong2024,kellogg2020,wood2021}. In online customer service centers, AI programs monitor calls, screen activities, and keystrokes to assess worker performance \cite{doellgast2021a,doellgast2023}. In large e-commerce firms like Amazon, algorithmic management uses wearable devices to track location and movements, creating high-resolution depictions of worker activity in the physical world \cite{crawford2021,delfanti2021}. AI management has been credited with increasing company profits and organizational efficiency \cite{sun2021,yaraghi2024}, but these gains often emerge through cost-cutting strategies and intensified worker discipline — for example, when AI-managed workers in gig economy sectors such as ride-hailing and delivery service drive longer hours for lower earnings \cite{bucher2021,delfanti2021,muldoon2023,rahman2021,wood2021}. Such conditions are fertile ground for worker backlash against AI management, which may take the form of gaming the system, disengaging from the platform, or organizing collective resistance \cite{anteby2018,doellgast2023,mohlmann2018}. 

Yet, experimental findings on how workers respond to AI-driven evaluation and compensation remain inconclusive — leaving us without clear behavioral insight into the consequences of AI management under controlled causal conditions. Indeed, experimental studies have produced contradictory evidence on how workers respond to AI management. Some suggest that AI management can reduce motivation and task engagement \cite{ranganathan2020,schlund2024}, as it is perceived to be reductionistic and lack the subjective capacities required for fair, respectful, or empathetic evaluation \cite{acikgoz2020,lee2018,dong2024,schlund2024}. In contrast, other studies report increased productivity under AI management \cite{bai2020,duani2024} when workers value the consistency and impartiality of AI, particularly when expecting biased treatment from human managers \cite{bigman2023,garvey2023,jago2022,yalcin2022}. These divergent findings reflect not only theoretical tensions but, perhaps more importantly, methodological challenges in designing ecologically valid simulations of AI management in a laboratory setting. 

Specifically, controlled experiments often struggle to simulate the following features of real-world AI management. \textbf{Lack of real-time monitoring}: Hypothetical scenarios rarely recreate the psychological pressure of being evaluated and observed in real time, a hallmark of AI management \cite{bucher2021,cameron2022,dong2024}. \textbf{Restricted autonomy in task execution}: Stripped-down experimental tasks can be too restrictive in terms of the discretion and flexibility that workers have in achieving their goals, which may confound reactions to AI management by amplifying perceptions of control \cite{mohlmann2018,rahman2021}. \textbf{Absence of motivational feedback loops}: One-shot tasks fail to capture how repeated cycles of AI evaluation and feedback affect workers' performance and motivation over time. \textbf{Absence of evaluation-based wages}: Hypothetical scenarios do not tie performance evaluations to actual contingent pay, limiting our ability to measure behavioral responses to wage reductions imposed by AI management. \textbf{Unfamiliar workplace}: Unlike real workers, who have a sense of their own competence, experimental participants are often assigned unfamiliar tasks that leave them without meaningful expectations or benchmarks against which to interpret AI evaluations \cite{barclay2017,critcher2009,sitzmann2010}. \textbf{Simulated AI}: Many vignette studies and experiments rely on pretended AI evaluations, which elicit responses to the idea of AI rather than to actual machine behavior; and the idea of AI can be more aversive than its real-world experience \cite{dong2024,jakesch2023,tong2021}. 

To overcome these challenges and isolate the behavioral consequences of AI-driven wage management, we developed a controlled yet ecologically rich experimental environment: the Iron Pickaxe Factory, a bespoke Minecraft game server. Minecraft is a widely adopted platform for training AI agents to imitate humans, navigate physical environments, and collaborate with human partners \cite{amresh2023,guss2021,hafner2025}. Its open-ended, programmable structure has also made it increasingly valuable for behavioral research, offering a flexible environment to study real-world cognition and behavior under controlled conditions \cite{bendell2024,peters2021,simon2022}. In our customized Minecraft workplace, participants repeatedly attempt to craft an iron pickaxe within a structured time window of 10 minutes, receiving a managerial evaluation and contingent wage after each attempt. After each evaluation, they report their fairness perceptions; and before each new attempt, they report their motivational state.

This experimental setup systematically reconstructs the key features of real-world AI management that laboratory studies struggle to capture. \textbf{Real-time monitoring}: Workers are supervised by a visible manager avatar throughout task execution (Fig.~1), while the system captures fine-grained behavioral data, including movement patterns, mining actions, item acquisition, and time-to-completion metrics. \textbf{Autonomy in task execution}: The iron pickaxe task allows multiple strategies for achieving the production goal, granting participants meaningful discretion in how they approach their work. \textbf{Dynamic motivational feedback loops}: Workers engage in repeated production cycles, each followed by a performance evaluation and contingent compensation, allowing us to track motivational dynamics over time. \textbf{Evaluation-based wages}: Workers' compensation is directly tied to evaluated performance, replicating the economic pressures of AI-managed labor and allowing us to observe behavioral responses to wage reductions imposed by AI versus human managers. \textbf{Familiar workplace}: Workers are Minecraft players reporting different levels of self-assessed skills, reflecting natural variation in worker ability and ensuring that they have meaningful internal benchmarks against which to interpret managerial evaluations. \textbf{Genuine AI management}: Our experiment deploys two functioning, specially trained AI managers that evaluate worker performance in real time, ensuring that participants respond to actual algorithmic behavior rather than imagined representations of AI.

\begin{figure}[h!]
    \centering
    \includegraphics[width=0.5\textwidth]{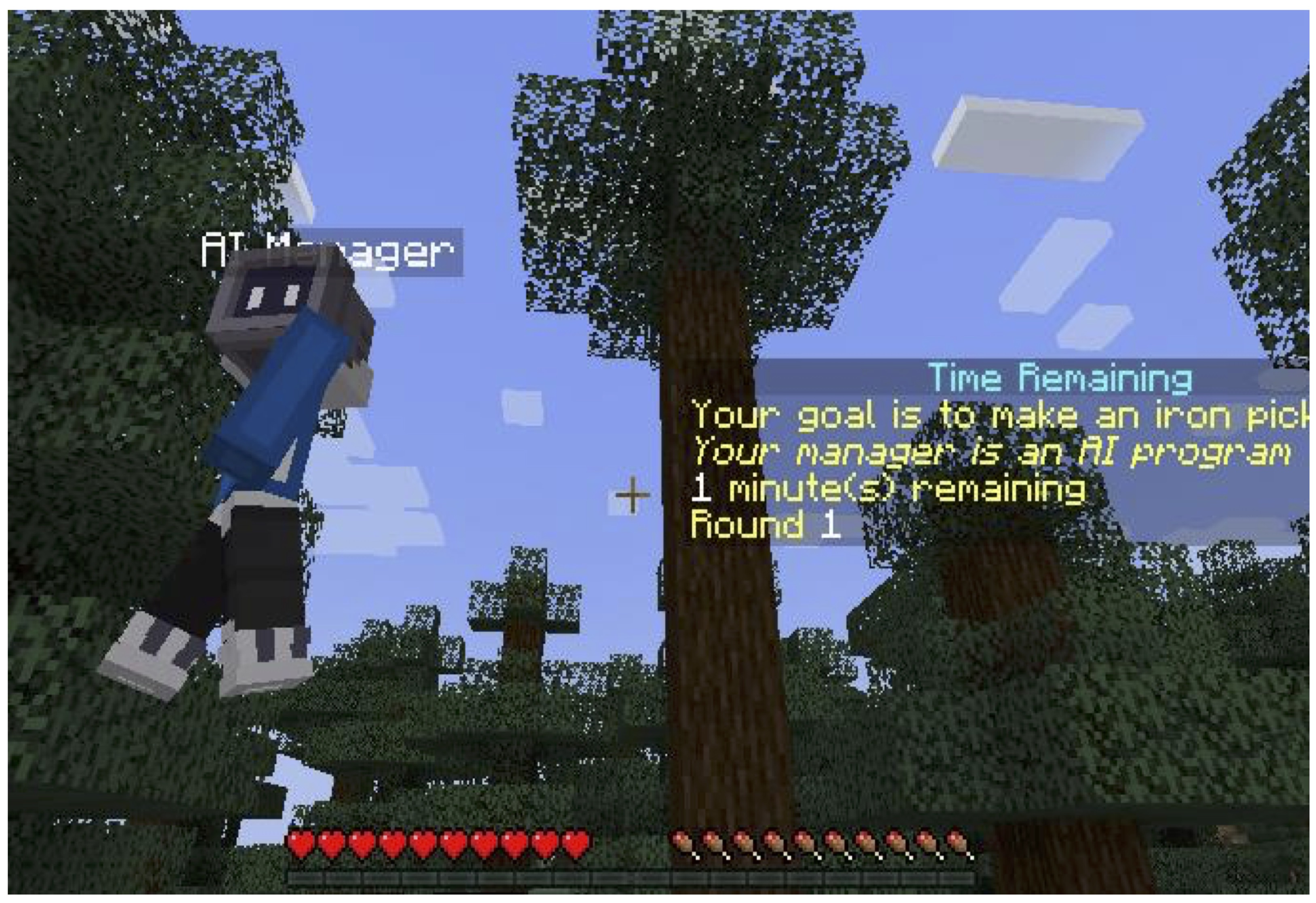} 
    \caption{\textbf{The experiment interface.} An example of the interface for workers under AI management. Workers complete the iron pickaxe task for three repeated rounds, each lasting 10 minutes. 
}
    \label{fig:sample-image} 
\end{figure}

With this experimental setup, we directly compare the effects of human versus AI management on worker evaluations, wages, fairness perceptions, and motivational dynamics. Here we show that an AI management system trained on human-defined evaluation principles evaluated worker performance more harshly than human managers, assigning lower scores than workers expect based on their self-assessment, and reduced wages by 40\% compared to human management. Critically though, fairness perceptions under human management tracked the discrepancy between self-assessment and evaluation, whereas they were less sensitive to this discrepancy under AI management. As a result, because motivation tracked perceived fairness regardless of manager type, AI-managed workers stayed motivated in spite of receiving lower wages, revealing a psychological blind spot to exploitation by AI. 

However, not all forms of AI management will produce this pattern. Indeed, we tested a different AI system (a decision-tree model trained on human assessments of annotated gameplay videos) which evaluated workers even more harshly, and decreased wages even more (60\% on average). However, this system led to a drop in fairness perceptions and failed to decouple fairness from performance evaluations, resulting in a drop in motivation. These findings suggest that while AI management can reduce compensation without backlash, this effect has boundary conditions: Some systems will cross a psychological threshold where management starts to feel like coercion, and compliance will give way to resistance.

\section{Results}

Participants joined our customized game server using their own Minecraft player account, and were randomly assigned to one of four management conditions. In the \textbf{Human} manager condition, workers were evaluated by another participant who had demonstrated top-tier performance during a pre-experimental training phase (see Methods). In the \textbf{AI-R} manager condition (R for 'Rules'), evaluations were provided by an algorithm trained on human-defined evaluation principles \cite{lawler1996}, based on the pace of advancement through the tech tree leading to an iron pickaxe. 

Two additional conditions tested further variations in evaluation sources. In the \textbf{Human + AI} condition, workers were evaluated by a human manager who received a recommendation from the AI-R system, allowing us to assess reactions to ostensibly human evaluations that are more closely aligned with AI-generated evaluations. In the \textbf{AI-T} manager condition (T for 'Trees'), evaluations came from an alternate AI system, a decision-tree model trained on the assessments of experienced crowdworkers watching annotated gameplay videos \cite{muldoon2023}. 

Prior to their first pickaxe attempt, all workers self-assessed their Minecraft competence as beginner, intermediate, or advanced. After each pickaxe attempt, managers evaluated workers using these same three categories, with associated wages of \textcent0 (beginner), \textcent20 (intermediate), and \textcent50 (advanced).

\textbf{Manager evaluations}. The modal evaluation by \textbf{Human} managers was ‘advanced’ (23\% beginner, 32\% intermediate, 46\% advanced), whereas the modal evaluation by \textbf{AI-R} was ‘intermediate’ (11\% beginner, 88\% intermediate, 1\% advanced), and \textbf{AI-T} was split between ‘beginner’ and ‘intermediate’ (47\% beginner, 52\% intermediate, 1\% advanced). As might be expected, evaluations in the \textbf{Human+AI} condition moved somewhat in between the Human and AI-R conditions (17\% beginner, 52\% intermediate, 31\% advanced). Since we recorded workers’ self-evaluations (8\% beginner, 61\% intermediate, 31\% advanced), we can measure discrepancies between self-evaluations and managerial evaluations across treatments (Fig.~2A). We conducted multilevel regression analyses to compare the propensity to downgrade or upgrade workers (compared to self-evaluations) across conditions. We find no credible evidence for a difference between the \textbf{Human} and \textbf{Human+AI} conditions (\(\beta = 0.18\), \(z = 0.35\), \(p = .73\), 95\% CI [\(-0.86\), \(1.22\)] for downgrading; \(\beta = -1.52\), \(z = -1.68\), \(p = .09\), 95\% CI [\(-3.29\), \(0.25\)] for upgrading). Both \textbf{AI-R} and \textbf{AI-T} were less likely than \textbf{Human} managers to upgrade workers (AI-R: \(\beta = -2.78\), \(z = -2.67\), \(p = .008\), 95\% CI [\(-4.83\), \(-0.74\)]; 
AI-T: \(\beta = -3.29\), \(z = -2.79\), \(p = .005\), 95\% CI [\(-5.61\), \(-0.98\)]). \textbf{AI-T} was also more likely to downgrade workers (\(\beta = 2.99\), \(z = 5.11\), \(p < .001\), 95\% CI [\(1.85\), \(4.14\)]), but there is no evidence that {AI-R} was (\(\beta = 0.64\), \(z = 1.21\), \(p = .23\), 95\% CI [\(-0.40\), \(1.69\)]).

\begin{figure*}[h!]
    \centering
    \includegraphics[width=\textwidth]{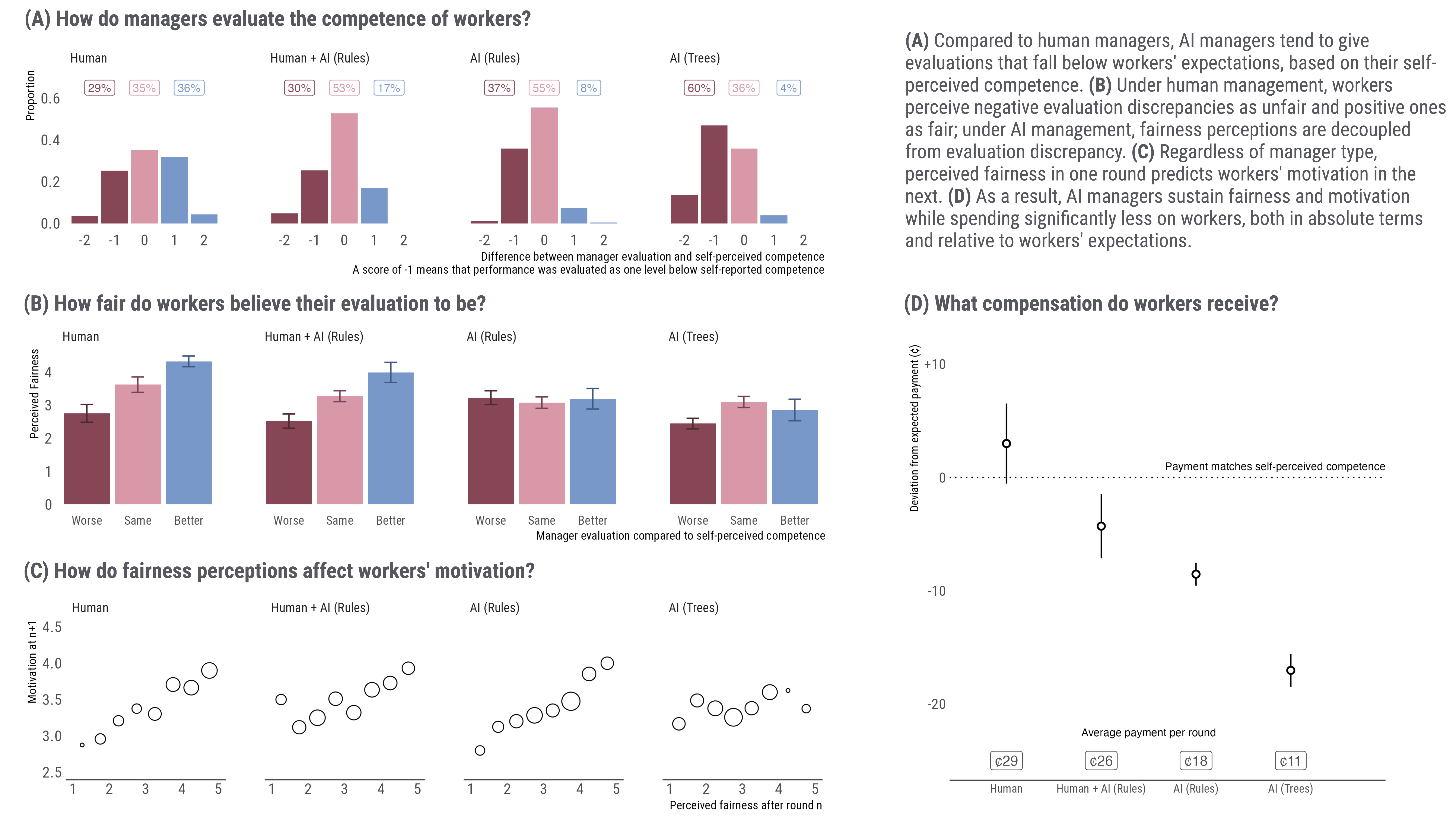} 
    \caption{\textbf{Summary of the main results (worker N = 382).} Both self-assessment and manager evaluation were implemented at three levels: beginner (\$0), intermediate (\$0.2), and advanced (\$0.5). Each level corresponds to a different bonus, as shown in the parentheses. Fairness scores aggregate procedural and distributive fairness; segmenting the two dimensions yielded similar results. 
}
    \label{fig:sample-image} 
\end{figure*}

\textbf{Worker wages}. AI managers paid workers less and deviated more from the payments workers could expect based on their self-assessed competence (see Fig.~2D). While there is no credible evidence of a difference in average wages between the \textbf{Human} and \textbf{Human+AI} conditions (average wages per round of \textcent29 and \textcent26, respectively; \(\beta = -0.20\), \(t = -1.84\), \(p = .07\), 95\% CI [\(-0.41\), \(0.01\)]), both \textbf{AI-R} and \textbf{AI-T} paid lower wages than \textbf{Human} managers (AI-R: \textcent18, a 40\% reduction, \(\beta = -0.68\), \(t = -6.35\), \(p < .001\), 95\% CI [\(-0.89\), \(-0.47\)]; AI-T: \textcent11, a 60\% reduction, \(\beta = -1.11\), \(t = -10.30\), \(p < .001\), 95\% CI [\(-1.32\), \(-0.89\)]). Compared to the \textbf{Human} manager condition, the deviation from the expected payment based on self-assessment was larger for \textbf{Human+AI} (\(\beta = -0.34\), \(t = -2.79\), \(p = .006\), 95\% CI [\(-0.59\), \(-0.10\)]), \textbf{AI-R} (\(\beta = -0.54\), \(t = -4.44\), \(p < .001\), 95\% CI [\(-0.79\), \(-0.30\)]), and \textbf{AI-T}(\(\beta = -0.95\), \(t = -7.66\), \(p < .001\), 95\% CI [\(-1.19\), \(-0.70\)]) conditions.

\textbf{Fairness perceptions}. Workers rated the fairness of the evaluation they received after each attempt at producing the pickaxe. These fairness perceptions are displayed in Fig.~2B, broken down by condition and by discrepancy between the evaluation and workers’ self-assessed competence. A multilevel regression detected a significant effect of this discrepancy: Workers felt treated more fairly when they were evaluated the same (\(\beta = 0.72\), \(t = 5.56\), \(p < .001\), 95\% CI [\(0.47\), \(0.98\)]) or better than they expected (\(\beta = 1.30\), \(t = 8.30\), \(p < .001\), 95\% CI [\(0.99\), \(1.61\)]), compared to when they were evaluated worse. This effect, however, was flattened in the \textbf{AI-R} condition compared to the \textbf{Human} condition, as detected by interaction effects (evaluated as same: \(\beta = -0.41\), \(t = -2.22\), \(p = .026\), 95\% CI [\(-0.78\), \(-0.05\)]; evaluated as better: \(\beta = -0.69\), \(t = -2.37\), \(p = .018\), 95\% CI [\(-1.27\), \(-0.12\)]). There was no credible evidence of such an interaction in the other conditions (all \(\beta\) between \(-0.47\) and \(0.09\), all \(p\) between .13 and .98). Overall, only the \textbf{AI-T} condition significantly lowered fairness perceptions compared to the \textbf{Human} condition (\(\beta = -0.41\), \(t = -2.72\), \(p = .007\), 95\% CI [\(-0.70\), \(-0.11\)]). There was no credible evidence of such an effect in the \textbf{Human+AI} condition (\(\beta = -0.31\), \(t = -1.91\), \(p = .057\), 95\% CI [\(-0.63\), \(0.01\)]), nor in the \textbf{AI-R} condition (\(\beta < 0.01\), \(t < 0.01\), \(p = .994\), 95\% CI [\(-0.33\), \(0.33\)]).

\textbf{Motivation}. Workers attempted the same task during three consecutive rounds, allowing us to examine whether fairness perceptions at round \(n\) predicted motivation at round \(n + 1\) in different management conditions. As shown in Fig.~2C, perceived fairness at round \(n\) significantly predicted worker motivation at round \(n + 1\) (\(\beta = 0.15\), \(t = 3.93\), \(p < .001\), 95\% CI [\(0.08\), \(0.23\)]), while the relation was not as strong in the AI-T condition (vs. the Human condition; \(\beta = -0.11\), \(t = -2.34\), \(p = .02\), 95\% CI [\(-0.21\), \(-0.02\)]). Accordingly, since fairness perceptions only decreased in the \textbf{AI-T} condition, we may expect that motivation only decreases in the \textbf{AI-T} condition, which is what we found (\(\beta = -0.14\), \(t = -2.17\), \(p = .03\), 95\% CI [\(-0.26\), \(-0.01\)]). There was no credible evidence of an impact on motivation in the \textbf{Human+AI} condition (\(\beta = -0.04\), \(t = -0.66\), \(p = .51\), 95\% CI [\(-0.17\), \(0.08\)]), nor in the \textbf{AI-R} condition (\(\beta = -0.07\), \(t = -1.18\), \(p = .24\), 95\% CI [\(-0.20\), \(0.05\)]).

\section{Discussion}

Our findings reveal that AI management can reduce wages without damaging worker morale. Workers evaluated by the AI system trained on human-defined rules (AI-R) received lower evaluations and substantially lower wages, but did not feel that they were treated unfairly, and maintained motivation over time. Key to this result is the fact that under human management, fairness perceptions closely tracked whether workers were rated above, below, or in line with their self-assessed skill. Under AI-R, this sensitivity to discrepancy was substantially reduced. One likely explanation for this flattened fairness response under AI-R is that algorithmic decisions are often perceived as more objective and less intentional \cite{bigman2023,duani2024,jago2022,lee2018,newman2020}, and thus provoke weaker emotional reactions: People may be less inclined to attribute disrespect to impersonal systems that operate without human discretion. This interpretation is supported by the results from the Human+AI condition. In that condition, where evaluations were shaped by AI-R recommendations but delivered by a human, fairness perceptions once again tracked discrepancies with self-assessed skills.

These findings should be interpreted in light of several limitations. First, although our main results hold for one type of AI manager trained on transparent, human-defined rules, they will not generalize to all AI systems. A second AI manager in our study (trained on crowdsourced evaluations of gameplay videos) issued even lower evaluations and wages, but triggered declines in both fairness perception and motivation. This may reflect the severity of the cuts, subtle differences in evaluative logic, or both. Second, our participant sample consisted of Minecraft players, who were predominantly male (70\%) and relatively young (mean age = 26, SD = 6.2), potentially limiting generalizability to more diverse labor populations. Third, while the Minecraft server allowed communication between human managers and workers, we disabled this feature to ensure comparability with AI-managed conditions. This reduces relevance to workplaces where manager–worker communication plays a central role. Fourth, workers could not communicate with each other, preventing any form of collective resistance or coordination. Finally, although the experiment lasted longer than typical lab tasks (approximately 40 minutes), it remains shorter than real-world employment relationships, and may not capture the long-term effects of AI versus human management.

Despite common perceptions of AI management as objective and lacking intentionality \cite{bigman2023,duani2024,jago2022,lee2018,newman2020}, our results suggest that algorithmic objectivity may not be inherently neutral when leveraged to normalize inequitable evaluation regimes. Notably, both AI manager models incorporated human inputs. The rule-based model was designed using human-defined metrics and was fitted to the distribution of human evaluation data. The decision-tree-based model was trained on crowd-sourced human evaluations of Minecraft gameplay videos. However, both AI managers deviated significantly from human assessments and workers’ self-perceived competence. These deviations were not random but systematically biased toward downgrades and underpayment. This pattern reflects broader legitimacy issues in AI training practices. In particular, outsourcing annotation labor to crowd workers without sufficient contextual understanding may result in misalignment with human intentions.

Our results are not good news. The very features that make AI systems appear impartial can also make them powerful instruments of silent exploitation, leading workers to accept downgraded evaluations and lower pay without protest. As AI management scales, its ability to neutralize outrage and cut wages becomes a structural threat to accountability. It will not be enough to evaluate AI systems based on efficiency or accuracy, especially in domains where there is no clear ground truth about worker performance. These systems must also be audited for their ability to bypass the social reactions that, while imperfect, have traditionally helped limit exploitative outcomes in human-managed work.

\section{Methods}

\textbf{Participants.} The gameplay video evaluation study (for training the decision-tree AI model; NO. C2021-13), the calibration study (for the decision-tree AI model; NO. A2022-16), and the official study (NO. A2023-02) all received ethics approval from the ethics committee at the Max Planck Institute for Human Development, and obtained informed consent from the participants recruited from Prolific. We had 382 workers (\(n = 94\) in the human manager condition; \(n = 95\) in the AI-advised human manager condition; \(n = 98\) in the rule-based AI condition; \(n = 95\) in the decision-tree-based AI condition) and 102 managers (\(n = 52\) in the human manager condition; \(n = 52\) in the AI-advised human manager condition). Simulation-based power analysis for multilevel regression models \cite{lafit2021} suggested that the sample size for workers was sufficient to detect the manager effect on fairness perception with higher than 90\% power at an alpha level of 0.05. Participants as workers had a mean age of 25.9 years (\(SD = 6.2\)), and were predominantly males (70.0\% males, 26.9\% females, and 3.4\% other), residing in the US (24.6\%), the UK (20.9\%), Poland (13.6\%), Portugal (7.9\%), and Germany (5.0\%). Upon completion, all participants received a basic participation fee of \$7.8 for the 40-minute study. Participants in the manager role received a bonus payment of \$0.5 each round, while others in the worker role received evaluation-compatible bonuses of \$0, \$0.2, or \$0.5 each round.

\textbf{Design and procedure.} Participants were scheduled to join the multi-player experiment through a prescreening survey hosted on Qualtrics, which included questions about their basic Minecraft knowledge, access to an active Minecraft account, and available timeslots to play Minecraft. Eligible participants were then organized into groups and invited to the official experiment at their available time through their Prolific ID. Through a Qualtrics survey, participants provided basic demographic information, self-assessed Minecraft competence, and open-ended text entries about how they expect AI and human managers would evaluate performance, respectively. They were then instructed to update their game to the latest version (1.19.4) and join the experiment using the institution-hosted server address.

After launching the experiment using a PC, all participants went through a practice session, familiarizing themselves with the iron pickaxe task in a training chamber. Those who passed the training session first were assigned to be managers, and other participants were randomly assigned to be workers under different managers.

In the manager role, human managers could switch between different workers and monitor their performance in real-time, while AI managers kept collecting workers’ behavioral data necessary for their evaluation. Both human and AI managers were prompted to evaluate each assigned worker’s performance at three levels: beginner, intermediate, and advanced. AI-advised human managers received a recommendation from the rule-based AI before each evaluation.

In the worker role, as shown in Fig.~1, participants were constantly reminded of their manager identity, the task of making an iron pickaxe, the present round, and remaining time in the round. Except for the different evaluation mechanisms, workers in the two human manager conditions were exposed to identical information (human avatar appearing labeled with “Your manager is a human player”) — similarly for workers in the two AI manager conditions (robot avatar appearing labeled with “Your manager is an AI program”). The procedure was identical in each round: participants worked on the task for 10 minutes, and indicated their motivation on four questions (enjoyment: “Did you enjoy this task?”; effort: “Did you put a lot of effort into this task?”; nervosity: “Did you feel pressured while doing this task?” [reverse-coded]; autonomy: “Did you feel that you have some choice about doing this task?”; adapted from \cite{deci1994,dong2024}; 1 = not at all to 5 = extremely). They then received their manager’s evaluation in the present round, and indicated their fairness perceptions (\(\alpha = 0.91\) across six items; adapted from \cite{colquitt2001}; procedural fairness: e.g., “Do you think the evaluation procedure was applied consistently to all workers?”, \(\alpha = 0.82\) across three items; distributive fairness: e.g., “Do you think the evaluation result was justified, given your performance?”, \(\alpha = 0.94\) across three items; 1 = not at all to 5 = extremely). At the end of the game, workers again provided open-ended text entries about assumed standards of their own manager’s evaluations.

\vspace{.2em}

\nocite{*}
\bibliography{aaai25}
\end{document}